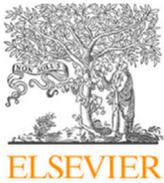

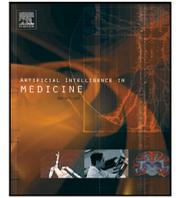

Research paper

# Domain generalization in deep learning based mass detection in mammography: A large-scale multi-center study

Lidia Garrucho [*], Kaisar Kushibar, Socayna Jouide, Oliver Diaz, Laura Igual, Karim Lekadir

*Artificial Intelligence in Medicine Lab (BCN-AIM), Faculty of Mathematics and Computer Science, University of Barcelona, Gran Via de les Corts Catalanes 585, Barcelona, 08007, Barcelona, Spain*

## ARTICLE INFO

## ABSTRACT

Computer-aided detection systems based on deep learning have shown great potential in breast cancer detection. However, the lack of domain generalization of artificial neural networks is an important obstacle to their deployment in changing clinical environments. In this study, we explored the domain generalization of deep learning methods for mass detection in digital mammography and analyzed in-depth the sources of domain shift in a large-scale multi-center setting. To this end, we compared the performance of eight *state-of-the-art* detection methods, including Transformer based models, trained in a single domain and tested in five unseen domains. Moreover, a single-source mass detection training pipeline was designed to improve the domain generalization without requiring images from the new domain. The results show that our workflow generalized better than *state-of-the-art* transfer learning based approaches in four out of five domains while reducing the domain shift caused by the different acquisition protocols and scanner manufacturers. Subsequently, an extensive analysis was performed to identify the covariate shifts with the greatest effects on detection performance, such as those due to differences in patient age, breast density, mass size, and mass malignancy. Ultimately, this comprehensive study provides key insights and best practices for future research on domain generalization in deep learning based breast cancer detection.

## 1. Introduction

Breast cancer is now the most common cancer worldwide, surpassing lung cancer for the first time in 2020 [1]. It is responsible for almost 30% of all cancers in women, and current trends show that its incidence is increasing [2]. In X-ray mammography (the gold standard imaging technique for early detection used in screening programs) breast cancer can be detected by identifying abnormalities in the breast structures, which may appear in the form of calcifications, architectural distortions, breast asymmetries, or masses. However, in breast cancer screening there is a high percentage of false-positives that may lead to unnecessary biopsies, along with a high rate of false negatives or missed cancers [3,4]. Overlooking or misinterpreting abnormalities found in mammograms are the most common reasons for missed breast cancers [5]. A recent large-scale study [6] compared the performance of an Artificial Intelligence (AI) system with the interpretation of 101 radiologists, concluding that the AI stand-alone approach achieved a cancer detection accuracy comparable to that of an average radiologist in the retrospective setting. Similarly, McKinney et al. [7] and Salim et al. [8] performed independent evaluations of several commercially available AI CADe systems and concluded that their diagnostic performance was enough to be further evaluated as an independent reader in prospective clinical trials. In contrast to these findings, a similar study [9] assessed the performance of AI algorithms from 126 teams and 44 different countries using mammograms from the United States and Sweden and concluded that the top-performing methods did not improve the radiologists' sensitivity. The contradictions in large-scale studies emphasize the importance of external validation in publications involving AI for breast cancer detection in mammography. Most AI detection methods are not tested for out-of-distribution (OOD) generalization using a different domain than the one used during training. The domain shift may lead to a substantial reduction in performance in different clinical settings — i.e. different scanner, imaging protocol, or patient cohort. To further increase the reliability and robustness of novel Computer Aided-Detection (CADe) methods there is an urgent need to study their generalization power and conduct external validation tests, as recommended by the FUTURE-AI guidelines [10]. Kim et al. [11] performed a meta-analysis of 516 published studies of the use of AI diagnostic analysis of medical images, fewer than 6% of which included external validation. Recently, Liu et al. [12] introduced the *medical algorithmic audit* to investigate and reduce the potential errors and harms caused by AI systems. In that study, the authors described all the potential domain shifts and highlighted the need for subgroup testing and analysis to increase the system domain generalization and trustworthiness.

In view of the above, Domain Generalization (DG) is an active research area that aims to improve the OOD generalization of AI solutions [13]. Most DG research in medical imaging has focused on the multi-source setting, which assumes that images from multiple domains are available in the training set. On the other hand, single-source domain generalization (SSDG) assumes that training images are homogeneous, coming from a single domain and lacking other domains during training. The goal of SSDG is to train a deep learning model to be robust against domain shifts using data from a single source domain. The SSDG setting is often more appropriate in medical imaging where public datasets are scarce and access to data is restricted. In this study, we investigated SSDG in the context of cross-domain breast cancer detection using digital mammography. In particular, we address breast mass detection, which is the most common pathology in public mammography image datasets. To the best of our knowledge, this the first study of SSDG in mammography. Our main goal was to develop a CADe system based on deep learning that is robust to domain shifts in digital mammography (Fig. 1).

In this paper, we address the DG problem in which both types of domain shifts are present. To sum up, the contributions of this paper are as follows:

- An extensive analysis of the mammograms in six different domains, highlighting the differences between domain and dataset shift and their potential effect on DG.
- A comparison of eight *state-of-the-art* detection methods, including Transformer-based architectures, fine-tuned for the task of mass detection in full-field digital mammograms using a single-source setting. The models' robustness was tested in five unseen domains, corresponding to different scanner manufacturers and datasets.
- The design of a SSDG training pipeline that boosts the breast mass detection performance and reduces the domain shift in unseen domains.
- Comparison of performance by mass and breast attributes, highlighting the potential biases of the proposed model.
- A study of the DG after using Transfer Learning on each unseen domain.

This study not only sheds more light on the domain generalization of deep learning, but also provides a comparison of *state-of-the-art* object detection methods in the challenging task of mass detection in mammography in different clinical environments. We also highlight the differences between domain and dataset shift in mammography and their possible effects on the detection performance. In the following section, we analyze the recent work on DG in medical imaging and mass detection in breast mammography.

## 2. Related work

### 2.1. Domain generalization in medical imaging

Samala et al. [14] studied the generalization error of a deep convolutional neural network fine-tuned for the task of classifying malignant and benign masses on mammograms. They aimed to balance the learning and memorization power of the network by varying the proportion of corrupted data in the training set. They concluded that training with noisy data, i.e. including 10% of corrupted labels, could increase the generalization error and improve the performance of transfer learning strategies. Wang et al. [15] outlined the inconsistencies in the performance of deep learning models in mammography classification in a study evaluating six deep learning architectures in a total of four datasets from different patient populations. Their results showed that the high performance obtained in the training dataset cannot be generalized to unseen external datasets, regardless of the model architecture, training technique or data labeling method. Recently, Li et al. [16] studied DG in lesion detection using a contrastive learning scheme to extract domain invariant features. The method was trained using mammograms from three different manufacturers and evaluated using mammograms from two unseen manufacturers, and showed great generalization power. For comparison with state-of-the-art generalization methods they only used the mean average precision (mAP) as the evaluation metric and no statistical significance tests were performed to confirm the improvement.

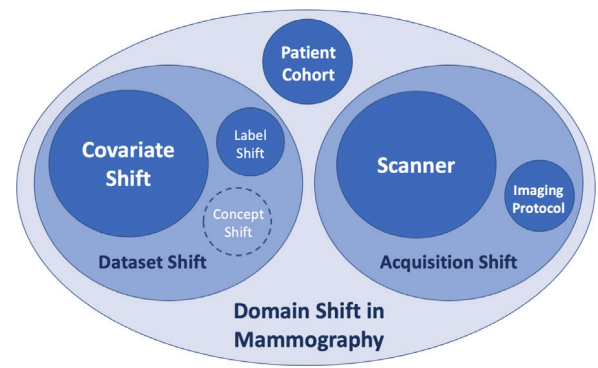

**Fig. 1.** Domain shift in mammography in a multi-center environment. *Covariate shift* — differences in the distribution of masses and breasts. *Label shift* — intra-observer variability of experts' annotations. *Acquisition shift* — different scanner manufacturers and imaging protocols. *Patient cohort* — differences in demography, geographic area, socioeconomic status, and patient comorbidities.

In other fields of medical imaging, like chest X-ray, prior work also found variable generalization performance of deep learning models in the presence of cross-institutional domain shift [17]. Cohen et al. [18] studied the generalization performance of chest X-ray prediction models when trained and tested on datasets from different institutions. The conclusion reached was that the shift present in the labels had a much higher impact on the generalization error than the domain shift in the images. In this study, we also discuss the differences between the domain shift, caused by the different image acquisition protocols, and the covariate shift, present in the data of each domain. Recently, Zhang et al. [19] benchmarked the performance of eight domain generalization techniques using multi-site clinical time series datasets and chest X-ray images. None of the DG methods achieved significant gains in OOD performance on the chest X-ray imaging data. In contrast to our study, they did not include intensity scale standardization nor any of the other single-source domain generalization techniques used in the current study. Moreover, a single classification architecture, a DenseNet-121, was trained using drastically down-sampled images.

In magnetic resonance imaging (MRI), Mårtensson et al. [20] examined the reliability of a deep learning model for clinical ODD data in the largest study to date on the effect of domain shift on deep learning models trained with MR images. They concluded that including more heterogeneous data from a wider range of scanners and protocols during training improved performance when using OOD data. In contrast, in our study, we focused on how to make models more robust when data from other institutions —i.e., domains are not available. Also in MRI, Ouyang et al. [21] proposed a causality-inspired data augmentation approach for single-source domain generalization for medical image segmentation and compared their method to other SSDG techniques, showing superior performance. In this study, we included some of the techniques tested in Ouyang et al. [21] to study their effectiveness in digital mammography.

In cardiac imaging, Zhang et al. [22] evaluated a deep stacked transformation data augmentation approach, named BigAug, with three different 3D segmentation tasks covering two medical imaging modalities (MRI and ultrasound) involving eight publicly available challenge datasets. In four different unseen domains, BigAug performed similarly to the two *state-of-the-art* methods. Finally, in digital pathology and histopathology, the domain shift effect for deep learning was studied in Thagaard et al. [23], Stacke et al. [24,25].

## 2.2. Mass detection in FFDM using deep learning

The detection and classification of masses in mammograms using deep learning has been covered in depth in the literature [26]. A wide variety of deep learning models have been developed to assist radiologists performing screening mammography. The models can be split into those that, during training, use as input a single mammogram [27–32] or multiple scans (generally both views of the same breast) [33–35], or patch-based approaches, using image patches [36–39].

Most methods in the literature report their performance in the same domain used for training while transfer learning is used afterwards to adapt the model to new domains. Here, we evaluated the generalization power of models trained in a single-source setting with and without DG techniques and tested their performance in unseen domains without using transfer learning. Additionally, we compared the best single-source DG model with transfer learning in five different domains.

Moreover, existing proposals in the literature employed a single well-known Convolutional Neural Network (CNN) architecture like Faster R-CNN [40] or YOLO [41] but the recent Transformer-based detection models [42–44] have not yet been fully explored. In the current study, we also included these novel Transformer-based detection models and evaluated their generalization power compared to other CNN detection methods.

### 2.2.1. Robustness of Transformer-based architectures

The OOD robustness of Transformer architectures has recently become a focus of studies since Transformers have established themselves in Computer Vision tasks [45–47], and specially since Visual Transformers (ViT) were introduced [48]. Most of these studies concluded that, due to the intrinsic properties of Transformers (which are mainly self-attention mechanisms and lack of strong inductive biases of convolutions), they outperform CNNs in terms of OOD robustness.

As an example, Zhang et al. (2021a) used the most popular data-shift datasets of ImageNet [49] and reported superior performance of the Transformer-based model, the DeiT [50], against a single variant of the popular Big Transfer (BiT) CNN-based model [51].

However, a more extensive analysis considering the most relevant variants of BiT and ViT concluded that Transformers are not more robust than CNN models, but are better calibrated. Pinto et al. [52] also questioned the superior robustness of Transformers attributed only due to their architecture components, e.g. self-attention mechanism and lack of inductive biases. They showed that the impact of pre-training is more important than the lack of self-attention, achieving better performance compared to Transformers with a CNN pre-trained with weakly supervised procedures using a large amount of data.

In summary, a good understanding of why self-attention mechanisms learn better representations in certain settings and how different pre-training strategies dramatically impact the downstream task is still lacking. In this study, we will compare the robustness of CNN-based versus Transformer-based object detection architectures trained using large-scale datasets and fine-tuned for the specific task of mass detection using a medium size digital mammography dataset (2,864 mammograms included in the training set).

### 2.2.2. Transfer learning in breast cancer detection

Transfer learning has been used in mammography breast cancer detection to adapt the model to new domains, mainly new scanners and imaging protocols [28,37,53,54]. Nevertheless, transfer learning presents two main drawbacks in medical imaging, namely data availability and catastrophic forgetting. Catastrophic forgetting [55] is a phenomenon of artificial neural networks that occurs when a model is trained sequentially on multiple tasks, abruptly forgetting previously learned information upon learning something new. When fine-tuning the model in a new domain, there is a risk of over-fitting the model to the new test set, which may have less diversity of masses than the original training dataset. Nevertheless, the transfer learning performance can only be assessed if data from the new domain are available, which is not always the case when developing a new CADe system.

## 3. Full-Field Digital Mammography (FFDM) datasets

Several open access X-ray mammography repositories can be found in the literature [56]. Here, three FFDM datasets were used to study the robustness and generalization of the selected methods in different domains: a subset of OMI-DB, also known as the OPTIMAM dataset [dataset] [57], INbreast [dataset] [58] and the Breast Cancer Digital Repository (BCDR) [dataset] [59], comprising a total of 4,352 FFDM (2,382 subjects) from six different domains including different scanner manufacturers and datasets. See Table 1 for a detailed list of the number of cases, mammograms, and annotated masses present in each domain.

### 3.1. OPTIMAM mammography image database

The OPTIMAM dataset [dataset] [57] is a shareable resource of digital mammography images from breast screening centers in the UK which includes DICOM images, experts' annotations and clinical observations (e.g., pathology reports). The database contains FFDM of women with screen-detected cancers and representative samples of normal and benign screening cases.

A subset of OPTIMAM, containing a total of 3,500 malignant and 500 benign cases, was used in this study. Each case in the dataset may contain different studies from the same patient. We made sure to split the training, validation and test sets by case and not by study. The two most common views of each breast were used as independent inputs: the medio-lateral oblique (MLO) and cranio-caudal (CC) view. The image matrix of the mammograms was $3328 \times 4084$ or $2560 \times 3328$ pixels depending on the vendor and the compression plate used for image acquisition.

OPTIMAM meets the requirements for a multi-center and multi-scanner study as it contains screenings from a total of three different centers and different scanner manufacturers. Among the different scanner manufacturers, only cases with annotated masses were selected. Four different domains were built from OPTIMAM: Hologic Inc., Siemens, GE and Philips (see Table 1), splitting the cases by scanner manufacturer.

### 3.2. INbreast dataset

INbreast mammograms [dataset] [58] were acquired from a single Portuguese center using a FFDM system, the MammoNovation from Siemens. Images, distributed in DICOM format, have a matrix of $3328 \times 4084$ or $2560 \times 3328$ pixels, depending on the compression plate used for image acquisition. This public database consists of a total of 115 cases with different lesion types including masses, calcifications, asymmetries and distortions. Out of 115 cases only 50 contain masses, including a total of 116 annotations. Most INbreast lesions are not biopsy-proven and the malignancy of the mass is classified based on the BI-RADS assessment categories [60]. It is common to group masses with BI-RADS $\in \{2,3\}$ as benign and masses with BI-RADS $\in \{4,5,6\}$ as malignant. INbreast was used as a single domain in this study.

**Table 1**

The total number of cases, mammogram images, and annotated masses in the six domains evaluated in this study. Each domain corresponds to different scanner manufacturers and databases.

| Dataset | OPTIMAM Hologic | OPTIMAM Siemens | OPTIMAM GE | OPTIMAM Philips | INbreastSiemens | BCDR |
|---|---|---|---|---|---|---|
| Cases | 1924 | 65 | 45 | 208 | 50 | 90 |
| Images | 3446 | 120 | 83 | 407 | 107 | 189 |
| Masses | 3603 | 126 | 85 | 419 | 116 | 199 |

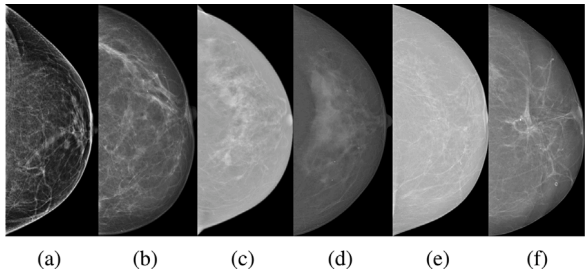

**Fig. 2.** Sample mammograms from the six domains: (a) OPTIMAM Hologic (b) OPTIMAM Siemens (c) OPTIMAM GE (d) OPTIMAM Philips (e) INbreast and (f) BCDR dataset.

### 3.3. Breast Cancer Digital Repository (BCDR)

The BCDR dataset [dataset] [59] is a public dataset from 2012, currently discontinued and available by request [61,62]. The dataset contains both digital (BCDR-DM) and film mammograms (BCDR-FM). In BCDR-DM, the dataset selected, a total of 90 subjects have biopsy proven mass lesion annotations. All images were supplied by the Faculty of Medicine-Centro Hospitalar São João, at University of Porto (FMUP-HSJ) and obtained using a MammoNovation Siemens FFDM scanner. Images have a matrix of 3328 × 4084 or 2560 × 3328 pixels, depending on the compression plate used for image acquisition, and are available only in 8-bit depth TIFF format. BCDR was used as a single domain in this study.

### 3.4. Domain shift in mammography

In Fig. 1, we introduce the different domain shifts present in digital mammography. The size of the circles in Fig. 1 is directly related to their importance in this study. In particular, we focused on the acquisition and the covariate shift.

The acquisition shift is given by the use of different scanner manufacturers and image acquisition protocols. As the information of differences among imaging protocols is not available, we focused on mitigating the acquisition shift given by different scanner manufacturers. Fig. 2 presents sample mammograms from each domain used in this study. The most notable differences among domains are the changes in intensity values and the contrast between the fibroglandular tissues and the adipose areas of the breast.

As well as the acquisition shift, medical imaging datasets suffer from additional *covariate shift* caused by the different data distributions among datasets. Covariate shift is difficult to avoid due to the scarcity of data and privacy constraints that limit the availability of large-scale medical imaging datasets for training. In mammography mass detection, the covariate shift is caused by differences in the masses – i.e. shape, size, malignancy and location – and the biological variations between patients — i.e. age and breast density.

Masses or nodules can appear anywhere in the breast, with different shapes and sizes, and may look benign or malignant. Moreover, other factors like breast density can increase the difficulty of mass detection. In high density breasts, there is a higher probability that the dense tissues (parenchyma) will occlude (or even simulate) masses and other breast lesions. For this reason, the overall sensitivity of mammography for breast cancer detection is reduced by more than 20% in dense breasts [63], even though women with dense breasts have a 4–6 fold increased risk of breast cancer than those with low density breasts [64–66].

Table 2 shows the total number of masses in each domain categorized by mass size, status, patient age and breast density. Age and breast density information was not available for all the domains. The breast densities are split by BI-RADS categories [67], *BI-RADS A* being almost entirely fatty breasts and *BI-RADS D* being extremely dense breasts. Overall, the INbreast and BCDR datasets have a much higher percentage of benign masses than the other four domains. The BCDR dataset has the largest covariate shift, with 55% of benign masses, 15% of masses measuring less than 5 millimeters in diameter and the youngest patient distribution. In our experiments, we show the impact of this covariate shift on the domain generalization error of the mass detection system.

The other domain shift sources in Fig. 1 have an indirect impact on the performance that is difficult to assess with the information available in the datasets. The patient cohort shift is related to the model fairness. Including datasets from different countries in the training phase would make the models fairer and more generalizable to this type of shift. In this study, we trained with data from the UK and tested in datasets from Portugal. Still, we do not have the information to measure if the performance gap is given by the different patient populations rather than the covariate shift. Moreover, the concept shift is ignored because the target is the same for all datasets — to detect suspicious masses.

## 4. Methodology

Our analysis was carried out in three stages. First, a total of eight *state-of-the-art* object detection methods pre-trained using the COCO dataset [68] were fine-tuned using a single domain for the downstream task of mass detection and tested on five unseen domains. Second, we selected the most robust method as the baseline and tested the generalization error after using different SSDG techniques in the training pipeline. Third, we tested the improvement in performance after fine-tuning for each unseen domain and compared it to the performance of the single-source setting. In the following sections, we describe the deep learning based object detection methods that were included in this analysis as well as the data preparation pipeline and the SSDG techniques that were used.

### 4.1. Object detection methods

In this section, we describe the eight *state-of-the-art* object detection methods compared in this study.

#### 4.1.1. Anchor-based detectors

Since the development of CNNs, object detection has been dominated by anchor-based detectors. These methods predict objects with predefined scales, aspect ratios and classes over every CNN feature location in a regular, dense sampling manner. Anchor-based methods are generally divided into one-stage and two-stage methods depending on the times the coordinates of the anchors are refined, affecting both performance and computational efficiency. Among the anchor-based methods, one of the most successful approaches both in computer vision and medical imaging is Faster R-CNN.

**Table 2**

Distribution of annotated masses in the different domains classified by mass status, mass size, patient age and breast density. Each column contains the percentage and the total number of masses in each category. N/A corresponds to missing or incomplete information.

| | OPTIMAM Hologic | OPTIMAM Siemens | OPTIMAM GE | OPTIMAM Philips | INbreast | BCDR |
|---|---|---|---|---|---|---|
| **Mass status** | | | | | | |
| Benign | 9% | 8% | 2% | 0 | 35% | 55% |
| Malignant | 91% | 92% | 98% | 100% | 65% | 45% |
| **Mass size** | | | | | | |
| < 5 mm | < 1% | 1% | 0 | < 1% | 2% | 15% |
| 5–10 mm | 19% | 23% | 19% | 21% | 11% | 24% |
| 10–15 mm | 30% | 33% | 25% | 33% | 24% | 14% |
| 15–20 mm | 22% | 21% | 19% | 19% | 13% | 8% |
| 20–30 mm | 21% | 19% | 28% | 17% | 18% | 14% |
| > 30 mm | 8% | 3% | 9% | 10% | 32% | 25% |
| **Age** | | | | | | |
| < 50 | 6% | 2% | 6% | 3% | N/A | 20% |
| 50–60 | 36% | 43% | 42% | 34% | N/A | 24% |
| 60–70 | 42% | 43% | 27% | 44% | N/A | 34% |
| > 70 | 16% | 13% | 25% | 19% | N/A | 22% |
| **Breast density** | | | | | | |
| BI-RADS A | 10% | 5% | 28% | 5% | 36% | 37% |
| BI-RADS B | 48% | 9% | 43% | 31% | 35% | 22% |
| BI-RADS C | 22% | 6% | 11% | 2% | 22% | 36% |
| BI-RADS D | 4% | N/A | 2% | < 1% | 7% | 5% |
| N/A | 16% | 80% | 16% | 62% | 0 | 0 |

**Faster R-CNN** [40] is a two-stage anchor-based method consisting of a separate region proposal network (RPN) and a region-wise prediction network (R-CNN). Since its publication, many studies on object detection have focused on improving its performance using different strategies — i.e. redesigning the architecture, including attention mechanisms, and modifying the training strategy. Agarwal et al. [32] used a Faster R-CNN for mass detection, also trained with OPTIMAM mammograms from a single scanner manufacturer.

### 4.1.2. Anchor-free detectors

Anchor-free detection methods became popular with the emergence of FPN [69] and Focal Loss [70]. These methods find objects present in the image without preset anchors, eliminating hyperparameters and increasing their generalization ability.

**Adaptative Training Sample Selection (ATTS)** [71], which is a new method to automatically select positive and negative training samples according to statistical characteristics of the object, was proposed to bridge the gap between anchor-based and anchor-free methods.

**Probabilistic Anchor Assignment (PAA)** [72] is a new anchor assignment strategy for use with single-stage detectors rather than the more common strategy of determining positive samples using Intersection-over-Union (IoU).

**VariofocalNet (VFNet)** [73] was designed to learn an IoU-Aware Classification Score (IACS) as a joint representation of object confidence and localization accuracy and perform a more accurate ranking of candidate detection bounding boxes. A new loss function, named Variofocal Loss, is used to train the dense object detector. Combining these two components and a box refinement branch, a new dense object detector is built based on FCOS+ATTS architecture.

**AutoAssign** [74] makes positive label assignment fully data-driven and appearance-aware, presenting a new sampling strategy to determine positive samples known as label assignment. The authors claim that AutoAssign can automatically adapt to different data distributions and achieve superior performance without any further adjustment.

**YOLOF** [75] revisits feature pyramid networks (FPN) for one-stage detectors. The authors claim that it achieves comparable results to RetinaNet [70] and DETR [42], and is faster.

### 4.1.3. Transformer-based detection models

**DEtection TRansformer (DETR)** [42] is a query-based set-prediction method to eliminate the need for many hand-designed components in object detection. DETR streamlines the training pipeline as a direct prediction problem, adopting an encoder–decoder architecture based on Transformers [76].

**Deformable DETR** [43] alleviates the slow convergence and limited feature spatial resolution of DETR. The limitation of transformer attention modules in processing image feature maps has been tackled by attending only to a small set of key sampling points around a reference.

### 4.2. Data preparation and training

The domain selected for training in the current study was the subset of OPTIMAM images from Hologic Inc. (a scanner manufacturer). The dataset was split into training, validation and test sets with 70%, 10% and 20% of cases respectively. This training dataset contains 1,924 cases with annotated masses and 2,864 mammograms. Mass status (either benign or malignant) and conspicuity (a measure of the radiologists' difficulty detecting the mass) were balanced between the training, validation and test sets.

The image preprocessing pipeline consists of cropping the images to the breast region – discarding the background – resizing them to 1333 × 800 pixels while maintaining the aspect ratio and, finally, normalizing to the default mean and standard used in the pre-trained setup. The only data augmentation used was random image flipping, both vertically and horizontally.

In this study, we use the MMDetection (v.2.13.0) PyTorch framework [77] and the pre-trained models available in their GitHub repository.[1] All models were pre-trained on the COCO dataset [68] using 1333 × 800 pixel image resolution and fine-tuned for the task of mass lesion detection in FFDM scans. A single GPU (24 GB NVIDIA GeForce RTX 3090) was used to fine-tune the models during a maximum of 50 epochs, using a batch size of 2 and adjusting the learning rate as recommended in the framework. The default learning rates in the MMdetection methods were adjusted to train with two samples in a single GPU instead of the default two samples in an 8 GPU setting (dividing the learning rate by 8). During fine-tuning, the epoch with the best mean Average Precision and a 50% bounding box overlap (bbox_mAP_50) with the validation set was selected as the best model. Most methods converged before epoch 20; only PAA and DETR needed more than 30 epochs to reach convergence. Additional settings used in the fine-tuning of the eight methods are shown in Table 3. In anchor-based detectors, such as Faster R-CNN, the most important hyper-parameters are the anchor box scales and ratios. The anchors should follow the aspect ratio and size distribution of the masses. For Faster R-CNN, we followed the recommendation in Agarwal et al. [32] for the same dataset and scanner manufacturer used in this study.

---

[1] https://github.com/open-mmlab/mmdetection

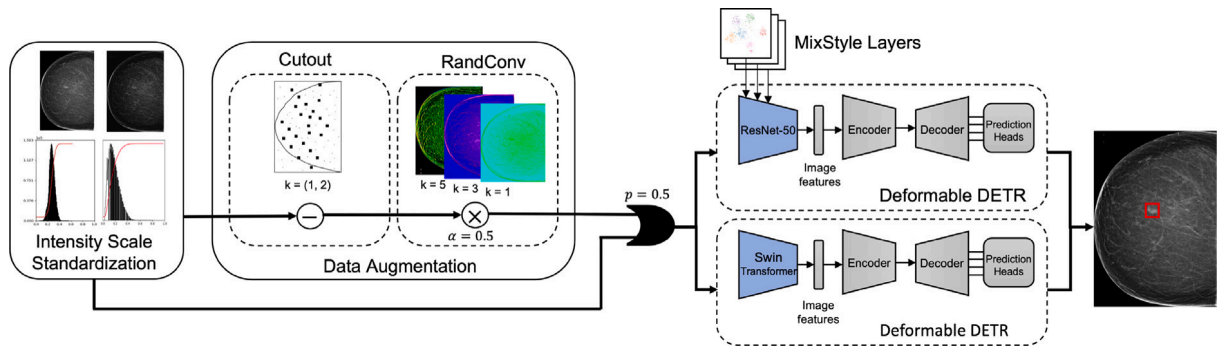

**Fig. 3.** Single-source domain generalization techniques applied during the training of the baseline model, a Deformable DETR.

**Table 3**
Training specifications of the models from the MMDetection framework. The backbone networks correspond to ResNet-50 (R-50) and ResNet-101 (R-101). The optimizer (opt) used was the model's default. The epoch where the model reached the convergence during fine-tuning is shown, along with the starting learning rate (LR), the steps used in the LR scheduler and additional settings specific to the model. All the models were initialized with pre-trained model weights from the COCO dataset available in the MMDetection GitHub repository.

| Method | Backbone | Opt | LR | epoch | Additional settings |
|---|---|---|---|---|---|
| ATSS | R-101 | SGD | 1.25e−03 | 9 | steps:[8, 11] |
| AutoAssign | R-50 | SGD | 1.25e−03 | 12 | steps:[8, 11,14] |
| Def DETR | R-50 | AdamW | 2.5e−05 | 10 | steps: [30]<br>iterative bbox refinement |
| DETR | R-50 | AdamW | 2.5e−05 | 48 | steps: [40] |
| Faster R-CNN | R-50-FPN | SGD | 2.5e−03 | 6 | steps:[8, 11]<br>scales:[0.1, 0.2, 0.5, 1.0, 2.0]<br>ratios:[0.5, 1.0, 2.0] |
| PAA | R-101-FPN | SGD | 1.25e−03 | 36 | steps:[28, 34], score voting |
| VFNet<br>DCN + MS train | R-50 | SGD | 1.25e−03 | 18 | steps:[16, 22] |
| YOLOF | R-50-C5 | SGD | 3.75e−03 | 9 | steps:[8, 11] |

### 4.3. Single Source Domain Generalization (SSDG)

Fig. 3 provides an overview of the different SSDG techniques tested in the training setup.

#### 4.3.1. Intensity Scale Standardization

Intensity, as well as texture, is a domain-dependent feature. CNNs are known to be susceptible to shifts in intensity [78]. There are two main approaches for removing the intensity shift among domains. First, using intensity-based data augmentation during training or, second, standardizing the intensities of the images before feeding them into the model — i.e. data harmonization. In this study we performed intensity scale standardization [79], which has been shown to greatly improve domain adaptation in medical imaging [80].

This technique was originally designed to standardize the intensity scales of MR images and ease the extraction of quantitative information. It is a two-step post-process method that aims to match similar intensities to similar tissue meaning. In the first step, the standardized histogram is learned from the training images, extracting the histogram landmarks. In the second step, the landmarks are used to linearly map the intensities of input images before feeding them into the network.

#### 4.3.2. Data augmentation methods for domain generalization

Two data augmentation methods were tested, namely Cutout [81] and RandConv [82]. Cutout helps the model to be robust to corruptions and missing features by deliberately removing square patches from training images at random locations. RandConv is a data augmentation strategy that generates images with random local textures but consistent shapes using linear filtering. The size of the convolution filter $k$ determines the smallest shape it can preserve. As an example, with $k = 2$, $2 \times 2$ random convolutions perturb shapes smaller than the filter size, which are considered local texture. Inspired by Augmix [83], the authors also propose blending the original image with the outputs of the RandConv layer via linear combination by a factor $\alpha$.

#### 4.3.3. Synthesizing novel domains using MixStyle

One way of increasing the diversity of source domains to improve OOD generalization is by synthesizing novel domains using only the training data. MixStyle [84] is a simple and versatile method inspired by style transfer. Capturing the style information from the bottom layers of a CNN and mixing styles of training instances results in novel domains that increase the diversity and hence the generalization of the trained model. The method mixes the feature statistics of two instances to synthesize new domains during the mini-batch training.

All the methods tested use a ResNet as the backbone to extract the image features. The authors of MixStyle recommend adding one

**Table 4**

Performance comparison of mass detection methods. The metrics correspond to the True Positive Rate (TPR) at 0.75 false positives per image (FPPI), the TPR 95% confidence interval (CI), and the AUC of the corresponding FROC curves. All the methods were fine-tuned using OPTIMAM Hologic manufacturer mammograms. The first column corresponds to the performance using the test set and the following ones evaluate the domain generalization performance using five unseen domains. The last column corresponds to the average AUC over the six different domains. The methods used were ATTS [71], AutoAssign [74], Deformable DETR [43], Faster R-CNN [40], DETR [42], PAA [72], VariofocalNet (VFNet) [73] and YOLOF [75]. The best-performing models are shown in bold.

| Method | OPTIMAM Hologic TPR (95% CI)/AUC | OPTIMAM Siemens TPR (95% CI)/AUC | OPTIMAM GE TPR (95% CI)/AUC | OPTIMAM Philips TPR (95% CI)/AUC | INbreast TPR (95% CI)/AUC | BCDR TPR (95% CI)/AUC | Avg AUC |
|---|---|---|---|---|---|---|---|
| ATSS | 0.931 (0.912,0.949) / 0.87 | 0.942 (0.901,0.983) / 0.90 | 0.678 (0.576,0.781) / 0.61 | 0.629 (0.582,0.677) / 0.55 | 0.716 (0.632,0.800) / 0.64 | 0.693 (0.626,0.759) / 0.57 | 0.69 |
| AutoAssign | **0.955 (0.940, 0.970) / 0.91** | 0.925 (0.880,0.970) / 0.89 | 0.715 (0.616,0.815) / 0.64 | 0.689 (0.644,0.735) / 0.58 | 0.721 (0.636,0.806) / 0.64 | 0.672 (0.605,0.740) / 0.60 | 0.71 |
| DETR | 0.942 (0.925,0.959) / 0.89 | 0.972 (0.943,1.001) / 0.93 | 0.681 (0.579,0.782) / 0.64 | 0.757 (0.714,0.800) / 0.69 | 0.833 (0.763,0.902) / 0.78 | 0.698 (0.633,0.764) / 0.63 | 0.76 |
| Faster R-CNN | 0.902 (0.880,0.924) / 0.84 | 0.889 (0.834,0.944) / 0.83 | 0.386 (0.280,0.491) / 0.38 | 0.319 (0.274,0.365) / 0.32 | 0.453 (0.359,0.548) / 0.45 | 0.450 (0.378,0.521) / 0.44 | 0.54 |
| PAA | 0.929 (0.911,0.948) / 0.88 | 0.917 (0.868,0.966) / 0.87 | 0.251 (0.155,0.346) / 0.22 | 0.338 (0.291,0.384) / 0.32 | 0.494 (0.400,0.589) / 0.47 | 0.576 (0.504,0.647) / 0.52 | 0.55 |
| VFNet | 0.943 (0.926,0.959) / 0.89 | 0.956 (0.921,0.991) / 0.91 | **0.777 (0.686, 0.868) / 0.71** | 0.652 (0.605,0.699) / 0.56 | 0.812 (0.740,0.884) / 0.70 | 0.672 (0.604,0.739) / 0.58 | 0.73 |
| YOLOF | 0.938 (0.921,0.956) / 0.89 | 0.908 (0.857,0.960) / 0.89 | 0.701 (0.602,0.800) / 0.65 | 0.683 (0.638,0.729) / 0.62 | 0.702 (0.616,0.787) / 0.62 | 0.625 (0.555,0.695) / 0.57 | 0.71 |

MixStyle layer after the first residual blocks of the ResNet, typically after block one, two and three, and testing which is the best configuration depending on the task.

### 4.3.4. Image feature extraction backbones

Another strategy to increase the model generalizability is to replace the image feature extraction backbone. By default, a convolutional backbone, the ResNet-50 or the ResNet-101, is used to extract the image features and add the MixStyle layers after the first residual blocks. To further test the power of Transformers, we replaced the convolutional backbone for a Transformer-based model, the Swin Transformer [44]. The Swin Transformer architecture was designed as a general-purpose computer vision backbone and out-performed the *state-of-the-art* architectures when it was published, showing the great potential of Transformers in the computer vision field.

### 4.4. Transfer learning on unseen domains

In a final experiment, we compared the transfer learning ability of the baseline model and the model trained using different SSDG strategies. To that end, the test datasets were split into training, validation and test sets using 80%, 5% and 15% of cases respectively. The models, previously fine-tuned using the OPTIMAM Hologic dataset, were fine-tuned again in the new domain. The fine-tuning settings were the same as in the previous experiments but convergence was reached before epoch 15 as the amount of data available for fine-tuning was small.

### 4.5. Evaluation metrics

In breast cancer mass detection, the True Positive Rate (TPR), also known as sensitivity or recall, is commonly used as the metric of reference to evaluate the performance of CADe systems [26]. The TPR penalizes the missed masses and rewards those that are detected. In commercially available CADe systems the TPR is typically reported as a range of (0.75, 0.85) false positives per image (FPPI) [28].

The area under the curve (AUC) of the Free-response Receiver Operating Characteristic (FROC) curve [85] was used to compare the methods. The AUC was computed by varying the confidence threshold of each bounding box in a range of FPPI $\in [0,1]$. A bounding box is a true positive (TP) when the Intersection-over-Union (IoU) of the prediction and the ground truth is greater than 10%, as recommended in [32]. Although 10% may seem very low for detection – an IoU of 50% is typically used in general Computer Vision – we evaluated the TPR versus the IoU threshold in the training dataset and confirmed that increasing the IoU by more than 10% had a negative impact on the TPR.

Following the recommendations of Demšar [86] for comparing multiple classifiers over multiple datasets, the Friedman test [87] was used to reject the null hypothesis. The Friedman test ranks the algorithms from best to worst for each dataset (domain) with respect to their performance (indicated by the AUC in this study). A post-hoc test is also needed to rank the algorithms from best to worst, comparing all classifiers with each other. In this case Demšar [86] suggests the Nemenyi test [88]. In the Nemenyi test, the performance of two classifiers is significantly different if the corresponding average ranks differ by at least the critical difference (CD).

## 5. Experiments and results

### 5.1. Performance comparison of mass detection models

Table 4 shows, for each domain, the performance of the methods in terms of TPR at 0.75 FPPI, 95% confidence intervals, and AUC. Fig. 4 contains the FROC curves for the six different domains. Comparing the FROC curves with the metrics in the table, we can see that a higher AUC correlates with the highest TPR at 0.75 FPPI.

All methods were able to detect masses in the source domain – the OPTIMAM Hologic – achieving a TPR higher than 90%. This good performance was maintained in the OPTIMAM Siemens domain, probably the dataset with the least shift from the source domain. The other four domains showed a significant drop in the AUC, with OPTIMAM Philips, OPTIMAM GE and BCDR being the most affected. Among the five unseen domains, the Transformer-based detection methods, DETR and Deformable DETR, were the most robust. The Deformable DETR model showed the greatest generalization power with an average AUC of 0.79. The methods that performed worst in terms of TPR and AUC were PAA and Faster R-CNN.

To further confirm the statistical difference between the tested methods, we ran a Friedman chi-square test using the AUC over all six domains. The test gave a $p$-value of $5.38e-06$, rejecting the null hypothesis and confirming that the methods are not equivalent and that their mean ranks are different.

### 5.2. Single-source domain generalization techniques

In this experiment, the most robust method, the Deformable DETR, was selected as the baseline. Then, the different single-source domain generalization techniques were added sequentially to the training pipeline. As shown in Fig. 3, Intensity Scale Standardization was applied prior to any data augmentation. Later, Cutout and RandConv were added for data augmentation with 0.5 probability.

Finally, two different strategies to extract image features were tested. First, three MixStyle layers were included in the ResNet-50 backbone. Second, the backbone was replaced by a Transformer-based model, the Swin Transformer. In Cutout data augmentation, we tried a variety of sizes for the patches $(1, 2, 3, 4, 5)$ removing 5%, 10% and 20% of total pixels. The best performance was obtained using a patch size smaller than 3 pixels (patches of 1 and 2 pixels) and removing 10% of the total pixels. That configuration did not miss small masses and important texture information. In RandConv data augmentation, we ran experiments using a combination of $k = (1,3,5)$ kernel sizes, $\alpha = (0.5, 0.7, 0.9)$ and $mix = (0, 0.5)$ mixing factor. The best results for RandConv data augmentation were obtained using $k = (1,3,5)$, combining the outputs with the input images using $\alpha = 0.5$ and without mixing.

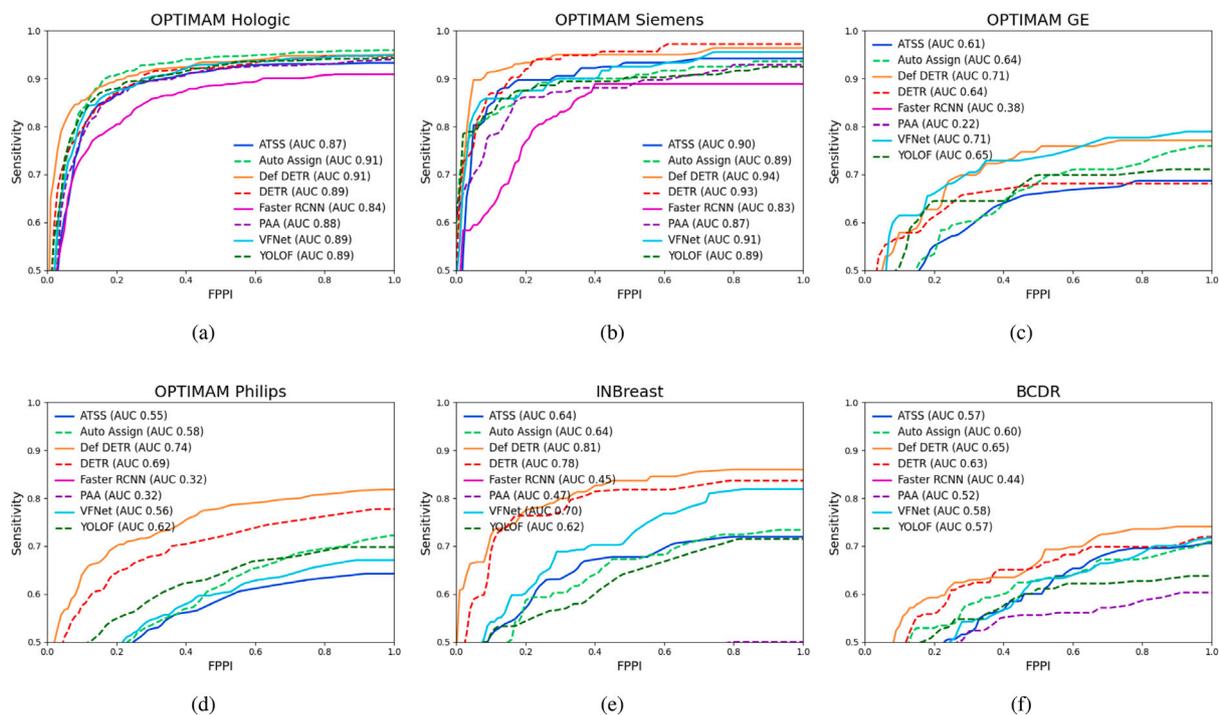

**Fig. 4.** Free-response Receiver Operating Characteristic (FROC) curves of the six different domains present in this study. The methods used were ATTS [71], AutoAssign [74], Deformable DETR [43], Faster R-CNN [40], DETR [42], PAA [72], VFNet [73] and YOLOF [75]. The eight mass detection methods were trained on the OPTIMAM (Hologic Inc.) dataset.

**Table 5**
Performance comparison of the baseline model, the Deformable DETR [43] and the model trained with different single-source domain generalization techniques: Intensity Scale Standardization (ISS) [79], Cutout (CO)[81], RandConv (RC) [82] and MixStyle (MS)[84], the Swin Transformer backbone (Swin Backbone) [44] and a combination of these. The metrics correspond to the True Positive Rate (TPR) at 0.75 false positives per image (FPPI), the TPR 95% confidence interval (CI), and the AUC of the corresponding FROC curves. All the methods were fine-tuned using OPTIMAM Hologic manufacturer mammograms. The first column corresponds to the performance using the test set and the following lines evaluate the domain generalization performance in the other five unseen domains. The models with the best performance are shown in bold.

| SSDG Technique | OPTIMAM Hologic<br>TPR (95% CI)/AUC | OPTIMAM Siemens<br>TPR (95% CI)/AUC | OPTIMAM GE<br>TPR (95% CI)/AUC | OPTIMAM Philips<br>TPR (95% CI)/AUC | INBreast<br>TPR (95% CI)/AUC | BCDR<br>TPR (95% CI)/AUC | Avg<br>AUC |
|---|---|---|---|---|---|---|---|
| Baseline | 0.948 (0.931,0.964) / 0.91 | 0.964 (0.933,0.995) / 0.94 | 0.771 (0.680,0.862) / 0.71 | 0.804 (0.765,0.842) / 0.74 | 0.858 (0.792,0.924) / 0.81 | 0.732 (0.668,0.795) / 0.65 | 0.79 |
| Intensity Scale Std (ISS) | 0.942 (0.925,0.958) / 0.91 | 0.931 (0.887,0.974) / 0.92 | 0.934 (0.880,0.987) / 0.89 | 0.896 (0.866,0.925) / 0.85 | 0.969 (0.939,0.998) / 0.94 | 0.729 (0.665,0.793) / 0.65 | 0.86 |
| Cutout (CO) | 0.948 (0.931,0.964) / 0.92 | 0.958 (0.926,0.991) / 0.94 | 0.838 (0.758,0.918) / 0.80 | 0.826 (0.790,0.863) / 0.78 | 0.890 (0.832,0.947) / 0.86 | 0.674 (0.606,0.741) / 0.63 | 0.82 |
| RandConv (RC) | 0.938 (0.921,0.956) / 0.90 | 0.958 (0.926,0.991) / 0.94 | 0.810 (0.721,0.898) / 0.75 | 0.789 (0.749,0.829) / 0.71 | 0.787 (0.711,0.863) / 0.73 | 0.656 (0.588,0.724) / 0.59 | 0.77 |
| MixStyle (MS) | 0.927 (0.909,0.946) / 0.90 | 0.950 (0.914,0.986) / 0.93 | 0.837 (0.757,0.917) / 0.79 | 0.830 (0.794,0.867) / 0.78 | 0.885 (0.826,0.943) / 0.84 | 0.677 (0.611,0.744) / 0.64 | 0.81 |
| Swin Backbone | **0.964 (0.951,0.977) / 0.93** | **0.969 (0.943,0.996) / 0.94** | 0.898 (0.832,0.963) / 0.85 | 0.917 (0.891,0.944) / 0.88 | 0.903 (0.849,0.958) / 0.88 | 0.716 (0.652,0.780) / 0.63 | 0.85 |
| ISS + CO | 0.938 (0.921,0.956) / 0.90 | 0.953 (0.918,0.988) / 0.92 | 0.982 (0.955,1.008) / 0.92 | 0.925 (0.899,0.950) / 0.87 | 0.991 (0.978,1.004) / 0.96 | 0.726 (0.663,0.790) / 0.65 | 0.87 |
| ISS + CO + Swin Backbone | **0.964 (0.951,0.977) / 0.93** | **0.969 (0.943,0.996) / 0.94** | **0.976 (0.942,1.010) / 0.93** | **0.929 (0.904,0.954) / 0.89** | 1.000 (1.000,1.000) / 0.99 | **0.765 (0.704,0.825) / 0.68** | **0.89** |
| ISS + CO + MS | 0.950 (0.934,0.965) / 0.91 | **0.969 (0.943,0.996) / 0.94** | 0.946 (0.898,0.994) / 0.89 | 0.902 (0.873,0.931) / 0.86 | **1.000 (1.000,1.000) / 0.99** | 0.726 (0.661,0.790) / 0.66 | **0.88** |
| ISS + CO + RC + MS | 0.939 (0.921,0.956) / 0.90 | 0.961 (0.930,0.992) / 0.93 | 0.946 (0.898,0.994) / 0.89 | 0.894 (0.864,0.924) / 0.84 | 0.982 (0.965,0.999) / 0.95 | 0.701 (0.635,0.767) / 0.62 | 0.86 |

Table 5 shows the performance of the models, for each domain, in terms of TPR at 0.75 FPPI, its 95% confidence intervals, and the AUC. The Deformable DETR trained with Intensity Scale Standardization showed the greatest gain in performance of all standalone methods tested, increasing the average AUC from 0.79 to 0.86, and boosting the TPR of the worst performing domains except BCDR. Similarly, the Swin Transformer backbone increased the average AUC from 0.79 to 0.85, being the best performing model in OPTIMAM Hologic and OPTIMAM Siemens domains. The third approach, adding Cutout data augmentation, improved the average AUC from 0.79 to 0.82. RandConv data augmentation seemed to downgrade the AUC in every domain except that of OPTIMAM GE. Last, MixStyle layers also helped to improve the performance on unseen domains with an average AUC of 0.81.

The last four models include a combination of the best performing SSDG methods. The models fine-tuned using the combination of Intensity Scale Standardization (ISS) and Cutout data augmentation (CO) gave the best results. The *ISS + CO + Swin Backbone* model boosted the average AUC by 3% in comparison to *ISS* alone, reaching 0.89 AUC. This model gave the best TPR and AUC in OPTIMAM GE, OPTIMAM

Philips and BCDR domains. The *ISS + CO + MS* model, adding MixStyle layers in the ResNet-50 backbone, boosted the average AUC to 0.88. This model was the best performing in INBreast domain, improving the baseline AUC from 0.81 to 0.99. Nevertheless, none of these SSDG techniques seemed to improve the mass detection performance in BCDR domain.

Other intensity-based data augmentations, such as histogram equalization and inverting intensity values, were also tested. However, the final performance deteriorated drastically and so the experiments are not included in this paper.

### 5.2.1. Computation complexity and performance

Intensity Scale Standardization (ISS) and Cutout (CO) data augmentation do not introduce additional hyper-parameters and so their computational burden is negligible. The major difference in computational complexity between the SSDG strategies is the number of parameters in the backbone architecture. The base architecture of Swin Transformer has a total of 88 M parameters while the ResNet-50 used in *ISS + CO + MS* strategy has 25.6 M. The inference times for the best-performing SSDG strategies are shown in Table 6. The GPU memory

**Table 6**

Computational complexity, GPU memory consumption, and training and inference times of the best-performing SSDG strategies. The image feature extraction backbone of *ISS + CO + MS* is a ResNet-50, while in *ISS + CO + Swin Backbone* is a base Swin Transformer. The batch size used to train the models was 1. Hardware: NVIDIA GeForce RTX 3090 24GB GPU and AMD EPYC 7272 12-Core Processor CPU.

| SSDG strategy | Backbone params | Training GPU memory/time | Inference GPU memory/time |
|---|---|---|---|
| *ISS + CO + MS* | 25.6 M | 17.5 GB - 37 min/epoch | 2.2 GB - 11.6 img/sec |
| *ISS + CO + Swin Backbone* | 88 M | 22.2 GB - 45 min/epoch | 5.1 GB - 3.1 img/sec |

**Table 7**

True Positive Rate, or sensitivity, of the **ISS + CO + MS** model for the different domains by mass status, size, patient age and breast density. N/A corresponds to missing or incomplete information. The subgroups with the worst performance are shown in bold.

| | OPTIMAM Hologic | OPTIMAM Siemens | OPTIMAM GE | OPTIMAM Philips | INbreast | BCDR |
|---|---|---|---|---|---|---|
| **Mass status** | | | | | | |
| Benign | 0.903 | 1 | 1 | N/A | 1 | **0.564** |
| Malignant | 0.949 | 0.948 | 0.934 | 0.897 | 1 | 0.888 |
| **Mass size** | | | | | | |
| < 5 mm | 0.333 | N/A | N/A | N/A | 1 | **0.276** |
| 5–10 mm | 0.954 | 0.931 | 0.938 | 0.861 | 1 | **0.542** |
| 10–15 mm | 0.922 | 0.951 | 0.905 | 0.942 | 1 | **0.750** |
| 15–20 mm | 0.954 | 0.963 | 1 | 0.913 | 1 | 0.938 |
| 20–30 mm | 0.950 | 1 | 0.917 | 0.944 | 1 | 0.964 |
| > 30 mm | 0.987 | 1 | 1 | **0.721** | 1 | 0.880 |
| **Age** | | | | | | |
| < 50 | 1 | 1 | **0.600** | 1 | N/A | 0.925 |
| 50–60 | 0.915 | 0.963 | 0.944 | 0.873 | N/A | **0.667** |
| 60–70 | 0.948 | 0.963 | 1 | 0.908 | N/A | **0.612** |
| > 70 | 0.968 | 0.875 | 0.952 | 0.899 | N/A | **0.705** |
| **Breast density** | | | | | | |
| BI-RADS A | 0.984 | 1 | 1 | 0.955 | 1 | **0.632** |
| BI-RADS B | 0.941 | 1 | 0.973 | 0.931 | 1 | 0.850 |
| BI-RADS C | 0.907 | 1 | **0.700** | 0.875 | 1 | 0.642 |
| BI-RADS D | 1 | 1 | 1 | **0.714** | 1 | **0.778** |
| N/A | 0.972 | 0.941 | 0.917 | 0.882 | 0 | 0 |

required during inference for the *ISS + CO + Swin Backbone* model doubles the one of *ISS + CO + MS*. In terms of inference speed, the *ISS + CO + MS* model processes 11.6 images per second while the throughput of the *ISS + CO + Swin Backbone* model is smaller, processing 3.1 images per second.

### 5.3. Detection performance by mass and breast attributes

Following the distribution of the datasets (Table 2), we tested the performance of the *ISS + CO + MS* model using the different domains by mass status, mass size, patient age and breast density. The sensitivity (TPR) values can be found in Table 7.

#### 5.3.1. Mass status

There is an imbalance between benign and malignant masses in the source domain: only 9% of the annotated masses in OPTIMAM Hologic are benign (see Table 2). When evaluated individually, the TPR for malignant and benign masses was 0.949 and 0.903, respectively. In OPTIMAM Siemens and OPTIMAM GE, the percentage of benign masses is 8% and 2% respectively, and all of them were detected. OPTIMAM Philips only contains malignant masses, which were detected with a TPR of 0.897. INbreast and BCDR are the domains with the most bias from the source domain in terms of mass status. In INbreast, 35% of masses are benign and all of them were detected, giving a TPR of 100%. On the other hand, in BCDR, 55% of the total masses are benign and only half of them were detected (TPR of 0.564).

#### 5.3.2. Mass size

To provide a better representation of the detected and missed masses by size, for each domain we illustrate the bounding box size distribution in Fig. 5. For comparison, Table 7 shows the performance by mass diameter, which is closely related to the width and height of the bounding box.

In OPTIMAM Hologic (Fig. 5(a)), most of the masses are between 5 and 25 mm in diameter. The range with the most samples (10–15 mm mass diameter) was also the range with the lowest sensitivity (0.922). Masses measuring less than 10 mm in diameter were correctly detected, confirming that the input size of the images after resizing was sufficient to detect small masses. Fewer than 1% of the masses in the dataset measured less than 5 mm and most of them were missed (0.333 TPR). Masses larger than 30 mm, 8% of the total, were correctly detected with a sensitivity of 0.987.

Like OPTIMAM Hologic, OPTIMAM Siemens (Fig. 5(b)) and OPTIMAM GE (Fig. 5(c)) do not have many masses larger than 30 mm in diameter and the missed masses were mostly between 5 and 20 mm. In OPTIMAM Philips (Fig. 5(d)), even though the mass size distribution is similar to the source domain, larger masses (> 25mm) were undetected (0.721 TPR). Some masses had a high height-to-width ratio compared to the other domains. In INbreast, all masses were detected, even those smaller than 5 mm in diameter. In BCDR, 15% of the masses are smaller than 5 mm in diameter and most of them were missed (0.276 TPR). The sensitivity in the range of 5 to 15 mm diameter was still low compared to the sensitivity of masses bigger than 15 mm in diameter.

#### 5.3.3. Age

In all OPTIMAM domains, most of the cases are from patients aged between 50 and 70 years old. However, in OPTIMAM Hologic, the performance in cases with patients younger than 50 and older than 70 years was better than that in the largest group. In OPTIMAM Siemens, the performance was worse in the group of patients older than 70 years, in contrast to OPTIMAM GE where the worst-performing group comprised patients younger than 50 years. Finally, in OPTIMAM Philips, all masses of patients younger than 50 were detected, and the performance was stable among the other groups. In BCDR, the cases are balanced among the four age groups, unlike in OPTIMAM GE, and performance was better in patients younger than 50 years compared to the other groups, which in turn showed worse but uniform performance. Information on age is not available for the INbreast dataset.

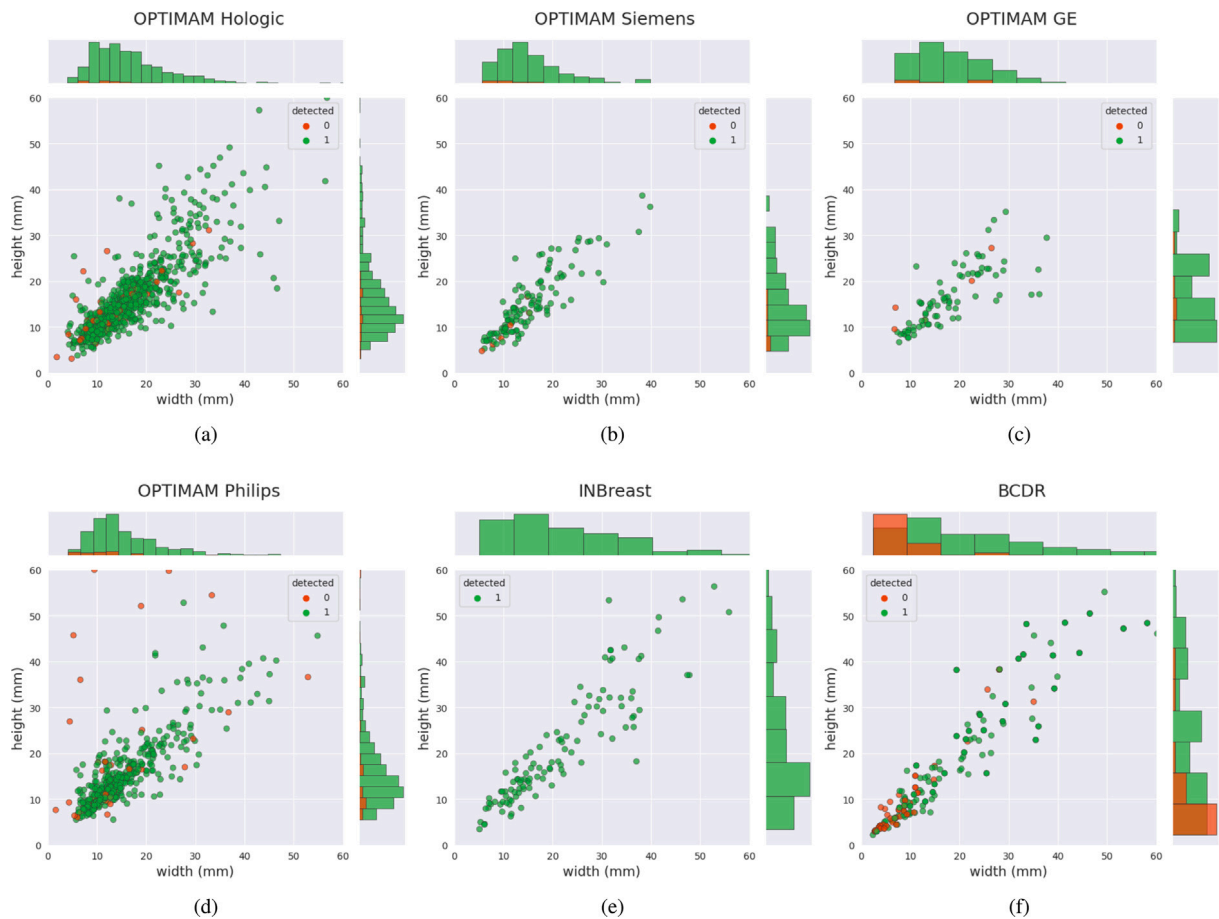

**Fig. 5.** Distribution of the detected (green) and missed (red) masses and their corresponding bounding box size in the different domains. The mass detections are from the best single-source domain generalization model trained, the Deformable DETR with Intensity Scale Standardization, Cutout data augmentation and MixStyle layers in the image feature extractor backbone (*ISS + CO + MS*). The bounding box width and height are in millimeters (mm). (For interpretation of the references to color in this figure legend, the reader is referred to the web version of this article.)

**Table 8**
Performance comparison of the baseline and the best-performing methods trained in a single-source setting with the models after applying transfer learning (TL) to each domain. The metrics correspond to the True Positive Rate (TPR), or sensitivity, at 0.75 false positives per image (FPPI), the TPR 95% confidence interval (CI), and the AUC of the corresponding FROC curves. The best-performing models are shown in bold.

| Method | OPTIMAM Siemens TPR (95% CI)/AUC | OPTIMAM GE TPR (95% CI)/AUC | OPTIMAM Philips TPR (95% CI)/AUC | INBreast TPR (95% CI)/AUC | BCDR TPR (95% CI)/AUC | Avg AUC |
|---|---|---|---|---|---|---|
| **Before TL** | | | | | | |
| Baseline | 0.886 (0.826, 0.945) / 0.86 | 0.757 (0.657, 0.858) / 0.70 | 0.751 (0.705, 0.797) / 0.68 | 0.840 (0.765, 0.916) / 0.79 | 0.724 (0.654, 0.794) / 0.65 | 0.75 |
| ISS + CO + MS | **0.908 (0.855, 0.961)** / **0.86** | 0.921 (0.859, 0.984) / 0.88 | 0.878 (0.844, 0.913) / 0.86 | **1.000 (1.000, 1.000)** / **0.98** | 0.728 (0.659, 0.798) / 0.66 | 0.84 |
| ISS + CO + Swin Backbone | 0.892 (0.834, 0.951) / 0.86 | **0.943 (0.887, 0.998)** / **0.89** | **0.878 (0.844, 0.913)** / **0.83** | **1.000 (1.000, 1.000)** / **0.98** | **0.770 (0.705, 0.835)** / **0.69** | **0.85** |
| **After TL** | | | | | | |
| Baseline | 0.854 (0.788, 0.920) / 0.82 | 0.786 (0.690, 0.882) / 0.76 | 0.841 (0.802, 0.879) / 0.78 | 0.921 (0.867, 0.976) / 0.86 | 0.831 (0.773, 0.889) / 0.76 | 0.80 |
| ISS + CO + MS | 0.870 (0.807, 0.933) / 0.84 | 0.907 (0.839, 0.975) / 0.86 | **1.000 (1.000, 1.000)** / **0.98** | **0.873 (0.821, 0.925)** / **0.80** | 0.84 |
| ISS + CO + Swin Backbone | **0.874 (0.811, 0.936)** / **0.85** | **0.929 (0.867, 0.990)** / **0.89** | **0.871 (0.835, 0.906)** / **0.82** | 0.994 (0.983, 1.006) / 0.97 | 0.842 (0.784, 0.900) / 0.78 | **0.86** |

### 5.3.4. Breast density

Breast density information is not available for all images in the OP-TIMAM dataset. In OPTIMAM Hologic, most cases are in the BI-RADS B category, but the performance was similar in all four categories. In OPTIMAM Siemens, only 20% of the cases have breast density information and all of them were correctly detected. In OPTIMAM GE, there was a drop in performance (0.70 TPR) in the BI-RADS C group while in OPTIMAM Philips there was a drop in the BI-RADS D group (0.714 TPR). Finally, the breast density distribution is similar in INBreast and BCDR. Nevertheless, in INBreast, all masses were correctly detected, while in BCDR, the sensitivity was only higher for the BI-RADS B category.

### 5.4. Transfer learning on unseen domains

In the next experiment, transfer learning was used to further adapt the models to every unseen domain. For this purpose, every domain was randomly split into training, validation and test sets, using 20% of the images for fine-tuning and 80% for testing. Table 8 shows the performance of the best SSDG strategies before and after transfer learning. The performances before transfer learning were computed again to allow fair comparison, as each domain was reduced to 20% for fine-tuning purposes. Before fine-tuning, the SSDG models continued to show the best performance, with gains of 9% and 10% over the baseline average AUC (0.75).

After fine-tuning using each domain, the baseline performance improved in all domains except for OPTIMAM Siemens, where the AUC dropped by 4%, from 0.86 to 0.82. The biggest improvements were seen in OPTIMAM Philips and BCDR, where the AUC improved by 10% and 9% respectively. OPTIMAM GE and INbreast improved their AUC by 6% and 7% each. After fine-tuning, both SSDG models only showed improvement in the BCDR domain. The *ISS + CO + MS* model increased the AUC in BCDR from 0.66 to 0.80, reaching a sensitivity of 0.873 at 0.75 FPPI. The AUC gain in BCDR for the *ISS + CO + Swin Backbone* model was smaller, from 0.69 to 0.78 with 0.842 sensitivity at 0.75 FPPI.

# 6. Discussion

In our first experiment, we compared the performance of eight detection methods fine-tuned for the task of mass detection in digital mammography. These methods comprised *state-of-the-art* anchor-based, anchor-free and Transformer-based detection methods. After evaluating their performance in five unseen domains, we concluded that Transformer-based methods were more robust to domain shifts in mammography datasets, the Deformable DETR being the best overall. As discussed in Section 2.2.1, the OOD robustness of Transformers has been highlighted in recent publications. Nevertheless, it would be misleading to conclude that their superior robustness is due to the intrinsic properties of Transformers — i.e. the self-attention mechanism and the lack of inductive biases. Regardless, we can conclude that, in our specific setting, Transformer-based methods learned better representations and generalized better in unseen domains than other detection methods. Thus, the Deformable DETR model trained was selected as the baseline for subsequent experiments.

In our second experiment, five different SSDG techniques were introduced in the training pipeline to improve OOD performance in unseen domains. ISS showed the greatest gain in performance among all stand-alone methods introduced, supporting the notion that deep learning detection models are strongly affected by the intensity distribution of the input images. In this regard, other intensity-based data augmentations, such as histogram equalization, were tested to further improve robustness, but were unsuccessful. We believe that the intensity alterations disturbed the data and added noise during training, worsening the final performance. This was confirmed when testing the RandConv method, which ultimately can be seen as an intensity-based data augmentation method, where training only with the augmented images worsened the performance, possibly because the intensity shifts looked unrealistic. The MixStyle layers and the Cutout data augmentation also helped improve the robustness among domains. In Cutout, patch sizes of 1 and 2 pixels were chosen, adding *salt and pepper* noise without the *salt* (bright pixels) to the data augmentation pipeline. For MixStyle, the best results were obtained by adding one layer after the first three residual blocks of the ResNet-50 used as the feature extractor. Replacing the ResNet-50 feature extraction backbone for a Transformer-based backbone, the Swin Transformer, also improved the baseline detection performance in all domains except for the BCDR domain. The combination of Intensity Scale Standardization and Cutout data augmentation gave the best results in all unseen domains in terms of average AUC and sensitivity for both backbones. Considering the performance difference of only 1% between the backbone strategies in terms of average AUC, the strategy selection will depend on the hardware capabilities and the processing speed requirements of the clinical setup.

To gain a better understanding of the model biases caused by different dataset shifts, we evaluated the detection performance by clinical and demographic variables such as mass status, mass size, breast density, and age. In the BCDR domain, there were big disparities in the performance by mass attributes. The performance dropped drastically for benign masses, which account for 55% of the masses in BCDR. In the source domain, the OPTIMAM Hologic dataset, benign masses represent only 9% of the total masses. Therefore, we could argue that the drop in performance in BCDR benign masses was caused by the class imbalance present in the source domain but, in contrast, in the INbreast dataset all benign masses (35% of the total) were detected. Moreover, all benign masses were detected in the other three unseen domains of the OPTIMAM dataset. Additionally, the model failed to detect small masses in BCDR, which is the domain with the highest proportion of masses smaller than 5 millimeters in diameter, accounting for 15% of the total. Regarding the mass size, the OPTIMAM dataset has few masses smaller than 5 millimeters in diameter and bigger than 30 millimeters. However, the performance in OPTIMAM domains was consistent over the different mass size groups except in OPTIMAM Philips, in which some of the masses missed in the detection have a higher height-to-width ratio than other domains (see Fig. 5(d)). Finally, we did not observe any correlation between the mass detection performance among different age groups and breast densities. Based on all these observations, we can conclude that the model seems to have a bias towards masses smaller than 5 millimeters in diameter and bounding boxes with a high height-to-with ratio, presumably because these samples were not representative in the training dataset (see Fig. 5(a)).

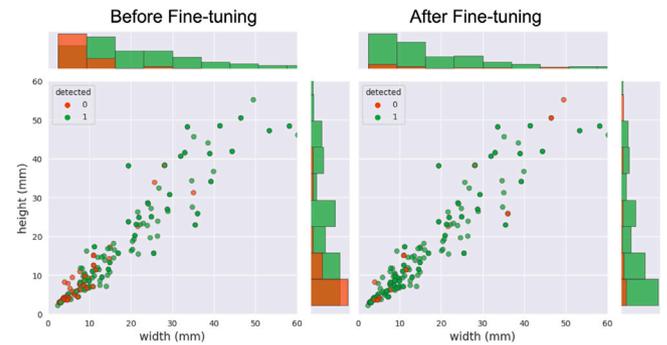

**Fig. 6.** Bounding box size distribution of BCDR dataset and the detected (green) and missed (red) masses of the *ISS + CO + MS* model before and after fine-tuning. The bounding box width and height are in millimeters (mm). (For interpretation of the references to color in this figure legend, the reader is referred to the web version of this article.)

As mentioned in Section 3.4, BCDR has the largest dataset shift in terms of mass size and mass status with respect to the training set, which may be why it was the worst-performing domain. In our last experiment, we found that Transfer Learning helped to mitigate the dataset shift in the BCDR domain. Fig. 6 shows that most of the small masses missed before fine-tuning were correctly detected after fine-tuning using 20% of the mammograms from BCDR. Inspection of the small masses missed before fine-tuning revealed that most of them contained small calcifications inside or surrounding the masses. Our reasoning is that this type of mass was not represented in the original training set and was not learned until fine-tuning in BCDR. However, the performance of the best-performing models, the *ISS + CO + MS* and the *ISS + CO + Swin Backbone*, in the other domains decreased after fine-tuning. This finding correlates with the risk of suffering catastrophic forgetting, one of the major limitations of applying Transfer Learning to a small dataset.

One limitation of this study was the imbalance in the training set in terms of mass and patient attributes. Adding more samples to the minority classes could help to evaluate better the detection performance and make a fairer comparison among subgroups. The minority classes in the training set were: benign masses, masses smaller than 5 millimeters in diameter, high height-to-width ratio bounding boxes, patients with high breast density and patients outside the age range of 50 to 70 years old. In addition, even if the Deformable DETR was the best performing method in our comparison, a very careful fine-tuning of the hyper-parameters of other detection models may bring the results closer to the Deformable DETR. However, such fine-grained searches for parameters are extremely time-consuming and limit the clinical applications of models with millions of trainable weights.

## 7. Conclusion

In this study, we evaluated different methods for mass detection in mammography using six different domains. Our experimental results showed that Transformer-based detection models were more robust to domain changes. Moreover, we highlighted the importance of using SSDG techniques to reduce the domain shift and improve performance in unseen clinical environments. The proposed training pipeline mitigated the domain shift present in four out of the five domains not seen during training. The results demonstrated that in one domain, the dataset shift caused by a higher proportion of small masses had a bigger impact than the domain shift caused by the acquisition pipeline. Additionally, we found that Transfer Learning helped to mitigate the dataset shift in that domain but decreased the performance in other domains. Transfer Learning is a powerful technique to mitigate the dataset shift, however, as shown here, it is not always successful and has to be applied carefully to avoid catastrophic forgetting. We believe that future work should focus on Continual Learning for AI in breast cancer detection. Continual Learning has great potential – either in a centralized or in a distributed manner – to allow CADe systems to avoid issues such as catastrophic forgetting, dataset shifts and demographic biases.

### Declaration of competing interest

The authors declare that they have no known competing financial interests or personal relationships that could have appeared to influence the work reported in this paper.

### Acknowledgments

This project received funding from the European Union's Horizon 2020 research and innovation programme under grant agreement No. 952103. A subset of the OPTIMAM database was obtained as part of a data-sharing agreement with the University of Barcelona in 2021. We are also grateful to Volpara Health (Dr Melissa Hill) for agreeing to share the breast density information available for the OPTIMAM subset used.